\documentclass[pra,aps,showpas,twocolumn,floatfix]{revtex4}
\usepackage{epsfig} \usepackage{graphics} \usepackage{bm}
\usepackage{amssymb} \usepackage{color}
\begin{document}
\title{Ground-state structure and stability of dipolar condensates in anisotropic traps }
\author{O.~Dutta\footnote{Corresponding author.
Email address: dutta@physics.arizona.edu} and P.~Meystre}
\affiliation{Department of Physics and College of Optical Sciences,
The University of Arizona, Tucson, AZ 85721, USA}
\date{\today}
\begin{abstract}
We study the Hartree ground state of a dipolar condensate of atoms
or molecules in an three-dimensional anisotropic geometry and at
$T=0$. We determine the stability of the condensate as a function
of the aspect ratios of the trap frequencies and of the dipolar
strength. We find numerically a rich phase space structure
characterized by various structures of the ground-state density
profile.

\end{abstract}

\maketitle

\section{Introduction}
The realization of Bose-Einstein condensation (BEC) in
low-density atomic vapors \cite{bec1, bec2} has led to an
explosion of experimental and theoretical research on the physics
of quantum-degenerate atomic and molecular systems. While much of
the work so far has concentrated on systems characterized by
$s$-wave two-body interactions, the recent demonstration of a
condensate of chromium atoms \cite{chbec} opens up the study of
gases that interact via long-range, anisotropic magnetic dipole
interactions. In a parallel development, it can be expected that
quantum degenerate samples of heteronuclear polar molecules will
soon be available through the use of Feshbach resonances
\cite{Kett, Jin1}, photoassociation \cite{DeM, HetDeM1}, or a
combination of the two. When in their vibrational ground state,
these molecules interact primarily via the electric dipole
potential, and they are expected to provide a fascinating new type
dipole-dominated condensates in the near future.

As a result of the anisotropy and long-range nature of the
dipole-dipole interaction, a number of novel phenomena have been
predicted to occur in low-density quantum-degenerate dipolar
atomic and molecular systems, both in conventional traps and in
optical lattices. An early study of the ground state of polar
condensates was presented in Ref. \cite{stabdp}, which determined
its stability diagram as a function of the number of atoms and
$s$-wave scattering length. It identified a stable structured
ground state for a specific range of parameters. At about the same
time, the effect of trap geometry on the stability of the
condensate was considered in Ref.~\cite{stabdp1} for a system
dominated by the dipole interaction. This was followed by the
prediction \cite{oplat} of the existence of a number of quantum
phases for dipolar bosons in optical lattices. Recent work
\cite{dvor1, dvor2} considers the structural phases of vortex
lattices in rotating dipolar Bose gases.

A novel feature of dipolar condensates, as compared to their
scalar cousins, is the appearance of a roton minimum in their
Bogoliubov spectrum. This feature was discussed in the context of
atomic condensates in Ref. \cite{rotmax}, which considered the
impact of the roton-maxon feature in the excitation spectrum and
the stability of pancake-shaped dipolar condensates. For this
particular geometry it was found that the excitation spectrum can
touch the zero-energy axis for a non-zero wave vector \cite{2dd},
which points to the instability of homogeneous condensates and the
onset of density modulations \cite{sup}. A roton minimum was also
found \cite{odell1} for the case of laser-induced dipolar
interactions in self-bound BECs with cylindrical symmetry.
Quasi-2D dipolar bosons with a density-modulated order parameter
were determined to be unstable within the mean-field theory
\cite{coop}, and cigar-shaped quasi-one dimensional condensates
were likewise found \cite{odell} to be dynamically unstable for
dipoles polarized along the axis of the cylindrical trap. The
stability of dipolar condensates in pancake traps was also
recently discussed in Ref. \cite{stabdp2}, which found the
appearance of biconcave condensates for certain values of the trap
aspect ratio and strength of the dipole interaction. From the
Bogoliubov excitation spectrum it was possible to attribute the
instability of the condensate under a broad range of conditions to
its azimuthal component.

Further building on these studies, the present note reports the
results of a detailed numerical analysis of the stability and
structure of the Hartree ground state of dipolar condensates
confined in anisotropic harmonic trap. We proceed by introducing
the trap frequencies $\omega_x, \omega_y$ and $\omega_z$,
respectively, in the $x$, $y$ and $z$ directions, and the
corresponding trap aspect ratios $\lambda_y = \omega_y/\omega_x$
and $\lambda_z=\omega_z/\omega_x$. Thus $\lambda_z =1$ corresponds
to a pancake trap, whereas $\lambda_z=0$ corresponds to a
cylindrical trap with free motion in $z$ direction. We further
assume for concreteness that an external field polarizes the
dipoles along the $y$ axis. The stability of the condensate is
then determined as a function of the trap aspect ratios and of an
effective dipolar interaction strength that is proportional to the
number of atoms or molecules in the condensate. Various ground
state structures of the condensate are identified in the stable
region of parameter space.

The remainder of this paper is organized as follows: Section II
introduces our model and comments on important aspects of our
numerical approach. Section III summarizes our results,
identifying up to five different types of possible ground states,
depending on the tightness of the trap and the particle number.
Finally, Section IV is a summary and conclusion.

\section{Formal development}
The dipole-dipole interaction between two particles
separated by a distance $r$ is
\begin{equation}
V_{\rm dd}(r) = g_{\rm dd} \frac{1 - 3 y^2/r^2}{r^3},
\end{equation}
where $g_{\rm dd}$ is the dipole-dipole interaction strength and
$\hat{y}$ is the polarization direction. For atoms with a
permanent magnetic dipole moment we have $g_{\rm dd} = \mu_0
\mu^2_m/4 \pi$ while for dipolar molecules $g_{\rm dd} = \mu^2_e/4
\pi \epsilon_0$, $\mu_m$ and $\mu_e$ being the magnetic moment of
the atoms and the electric dipole moment of the molecules,
respectively.

Within the mean-field approximation, the condensate order
parameter $\phi(r)$ satisfies the Gross-Pitaevskii (GP) equation
\begin{eqnarray}\label{gpeq}
    E \phi(r) & = &  \left [ H_0  + g \left | \phi^2(r)
    \right |  \right .   \nonumber\\
    & +  &  \left .  N \int V_{\rm dd}(r - r')
    \left | \phi(r') \right |^2  d^3r' \right ] \phi(r) ,
    \end{eqnarray}
where
\begin{equation}
H_0 = -\frac{\hbar^2}{2 m} \nabla^2 + \frac{1}{2} m \omega_x^2 \left
( x^2 + \lambda^2_y y^2 + \lambda^2_z z^2 \right )
\end{equation}
is the sum of the kinetic energy and the trapping potential and
$N$ denotes the number of particles in the condensate. The second
term on the right-hand side of Eq. (\ref{gpeq}) is the contact
interaction, $g=4\pi\hbar^2 a/m$ being proportional to the
$s$-wave scattering strength $a$, and the third term describes the
effects of the nonlocal dipole-dipole interaction. For dipole
interaction dominated systems, $g$ is small compared to $V_{\rm
dd}$. This is the case that we consider here, and in the following
we neglect the $s$-wave scattering term altogether.

For convenience we introduce the dimensionless parameter
    \begin{equation}
    D=N g_{\rm dd} m/(\ell_x \hbar^2)
    \end{equation}
that measures the effective strength of the dipole-dipole
interaction, where the oscillator length $\ell_x=\sqrt{\hbar/(m
\omega_x)}$. The condensate ground state is then determined
numerically by solving the Gross-Pitaevskii equation (\ref{gpeq})
for imaginary times. The term involving the dipole interaction
energy is calculated using the convolution theorem,
\begin{eqnarray}
    &&\int V_{\rm dd}(r - r') | \phi(r') |^2  d^3r'=
    \nonumber \\
    &&{\cal F}^{-1}
    \left \{{\cal F} \left [ V_{\rm dd}(r) \right ] * {\cal F} \left
    [ | \phi(r) |^2 \right ] \right \}, \nonumber
\end{eqnarray}
where ${\cal F}$ and ${\cal F}^{-1}$ stand for Fourier transform
and inverse Fourier transform, respectively. The dipole-dipole
interaction is calculated analytically in momentum space as
\cite{stabdp},
\begin{equation}
{\cal F}\left [ V_{\rm dd}(r) \right ]= \frac{4\pi}{3} \left [ 3
\frac{k^2_y}{k^2_x+k^2_y+k^2_z} - 1 \right ],
\end{equation}
$k_x, k_y, k_z$ are the momentum components in $x, y, z$
direction.

The initial order parameter was chosen randomly, and the stability
diagram was generated for each pair of parameters $(\lambda_y,
\lambda_z)$ by increasing the effective dipolar strength $D$ until
a critical value $D_{\rm cr}$ above which the condensate
collapses. Because of the random initial condition this value
varies slightly from run to run. The plotted results show the
average over 100 realizations of the initial wave function, the
error bars indicating the maximum deviation from of $D_{\rm cr}$
from its mean $\bar{D}_{\rm cr}$. This approach typically resulted
in numerical uncertainties similar to those of Ref. \cite{dip}.

\section{results}
\begin{figure}[ht]
\begin{center}
\epsfig{file=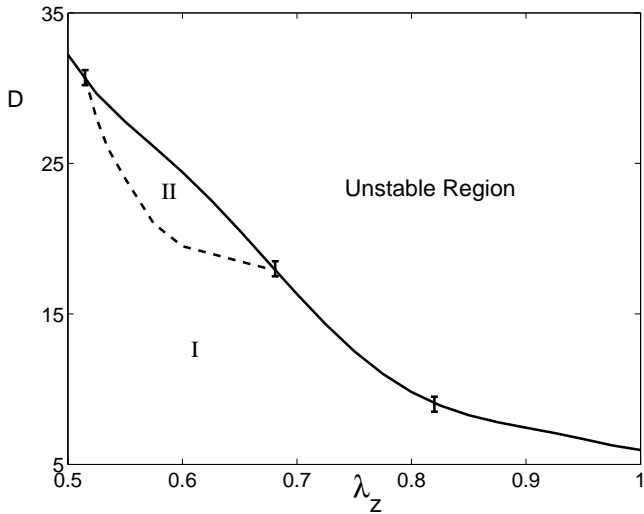,width=8.6cm} \caption{\label{fig1} $(\lambda_z,
D)$ stability diagram of a dipolar condensate in an anisotropic
trap for $\lambda_y=5$. The condensate is unstable in the region
above the solid line. The dashed line is the boundary between a
``structureless gaussian" and a double-peaked ground-state density
profile. The error bars give an indication of the accuracy of the
numerical simulations.}
\end{center}
\end{figure}

A good starting point for the discussion of our results is
the observation that in the case of a cylindrical trap,
$\lambda_z=0$, we found no stable structured condensate ground
state. (By structured profiles, we mean profiles that are not
simple gaussians.) In particular, solutions exhibiting density
modulations along the $z$-axis were found to be unstable. Moving
then to the case of a pancake trap by keeping $\lambda_y$ fixed
but increasing $\lambda_z$ from $0$ to $1$, we found for
$\lambda_y \gtrsim 4$ the appearance of a small parameter region
where the stable ground state is characterized by a structured
density profile, the domain of stability of this structured
solution increasing with $\lambda_y$. A $\{\lambda_z - D\}$ phase
space stability diagram typical of this regime is shown in
Fig.~\ref{fig1} for $\lambda_y=5$. In this figure, region $I$ is
characterized by the existence of a usual condensate with its
familiar, structureless gaussian-like density profile. As
$\lambda_z$ is increased, the condensate becomes unstable for
decreasing values of the effective dipole interaction strength
$D$, or alternatively of the particle number $N$. For $.525 <
\lambda_z < .7$, though, the ground state changes from a
gaussian-shaped to a double-peaked density profile (domain II in
the figure), before the system becomes unstable.

Figures \ref{fig4} and \ref{fig5} show surface plots and
corresponding 3-D renditions of density profiles typical of the
various situations encountered in our study. Figures \ref{fig4}a
and \ref{fig5}a are illustrative of the present case. The
appearance of two density peaks away from the center of the trap
results from the interplay between the repulsive nature of the
dipoles in a plane transverse to its polarization direction, the
$(x-z)$ plane, and the confining potential.

\begin{figure}[ht]
\begin{center}
\epsfig{file=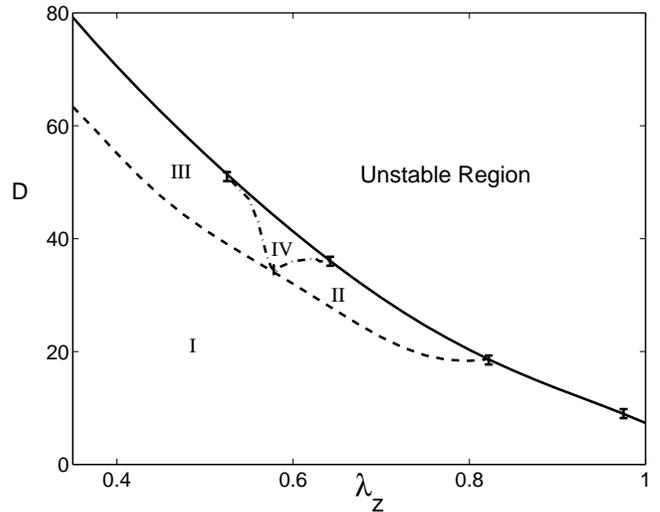,width=8.6cm} \caption{\label{fig2} Stability
diagram of a dipolar condensate in an anisotropic trap with
$\lambda_y=5.5$, as a function of the dipolar strength $D$ and
aspect ratio $\lambda_z$. The region above the solid line
corresponds to unstable solutions of Eq.(\ref{gpeq}). The dashed
line is the boundary between the structured and the standard
ground-state density profiles. The dot-dashed line marks the
boundary between the domains with quadruple-peaked and
double-peaked ground-state density profiles.}
\end{center}
\end{figure}

Increasing the tightness of the trap along the polarization
direction $y$, i.e, increasing $\lambda_y$, results in the
emergence of additional types of structured ground states. One
such case is illustrated in Fig.~\ref{fig2}, which is for
$\lambda_z=5.5$. For small values of $\lambda_z$, i.e., a weak
trapping potential along the $z$-direction, we observe the
appearance of a domain (region III in the figure) characterized by
a double-peaked ground state with the maximum density along the
$z$ direction and a gaussian-like density in $x$ direction. This
type of double-peaked structure along the weak trapping axis was
first predicted in Ref. \cite{dpeak}. Typical density profiles in
this region resemble those in Figs. 4a and 5a, but with a rotation
by 90 degrees in the $(x,z)$-plane. The regions II and III are
separated by a small additional domain IV characterized by a
ground-state distribution with a quadruple-peaked structure as
illustrated in Figs. 4b and 5b, as might be expected. In general,
these characteristics of the ground state density profile persist
until $\lambda_y \approx 6.5$.

\begin{figure}[ht]
\begin{center}
\epsfig{file=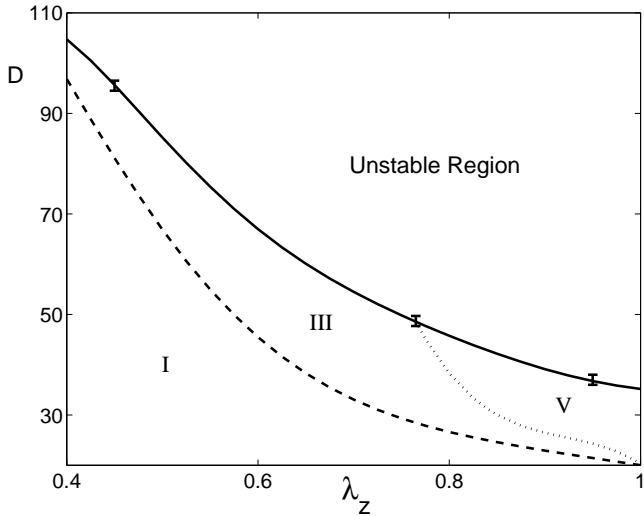,width=8.6cm} \caption{\label{fig3} Stability
diagram of a dipolar condensate in an anisotropic trap with
$\lambda_y=7$, as a function of the dipolar strength $D$ and the
aspect ratio $\lambda_z$. The region above the solid line is
characterized by unstable ground-state solutions of the mean-field
equation (\ref{gpeq}). The dashed line marks the boundary between
a structured and a gaussian-like ground-state density profile. The
dotted line is the boundary between the region with ring-like and
two-peaked condensate in $(x-z)$ plane.}
\end{center}
\end{figure}

\begin{figure*}[ht]
\begin{center}
\epsfig{file=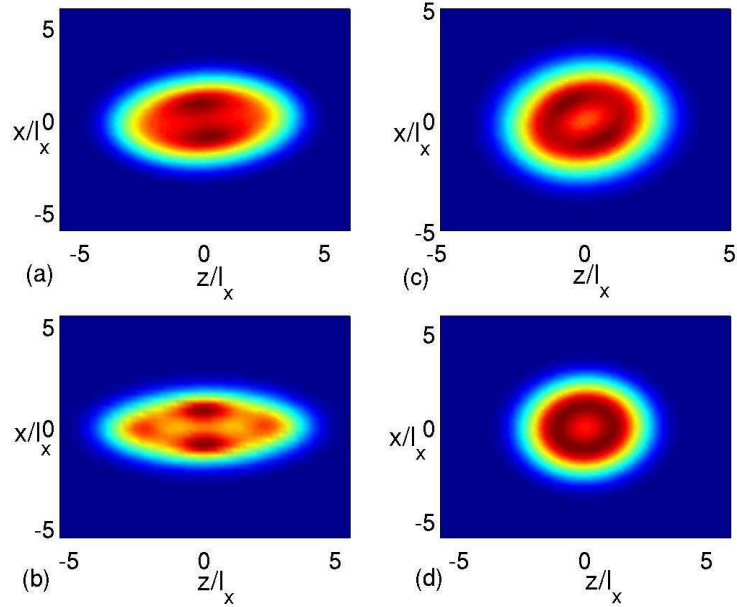,width=11.5cm} \caption{\label{fig4}
Two-dimensional surface plots of the structured ground-state
density profiles typical of various stable domains: (a)
Double-peaked density profile in the ($x-z$)-plane for region
$II$, for the parameters $\lambda_y=5$, $\lambda_z=.6$ and $D=23$.
The points of maximum density are away from the trap center and
along the $x$-direction. In region (III) the shape of the
condensate is similar, but with maximum density along $z$
direction. (b) Typical quadruple-peaked density profile
characteristic of region IV. Here $\lambda_y=5.5$,
$\lambda_z=.575$ and $D=42$ (c) Stable ground state solution in
region $V$ with $\lambda_z<1$. The density is higher and modulated
on a radius away from the trap center in the $x-z$ plane. In this
example $\lambda_y=7$, $\lambda_z=.85$ and $D=40$. (d) Density
profile typical of the domain V. The condensate is biconcave with
maximum density along a constant radius from the trap center. In
this example,\ $\lambda_y=7$ and $D=32$.}
\end{center}
\end{figure*}

\begin{figure*}[ht]
\begin{center}
\epsfig{file=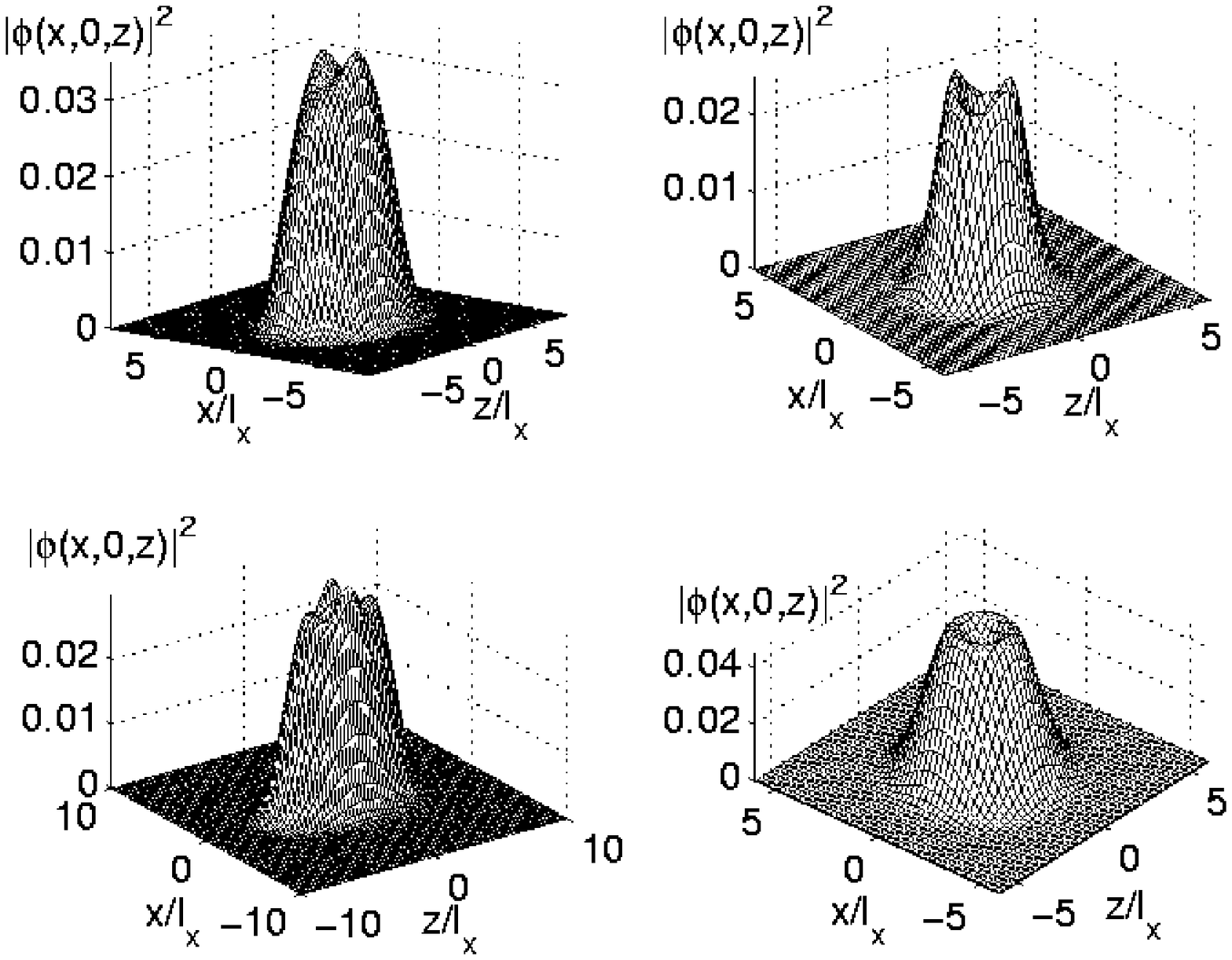,width=11.cm} \caption{\label{fig5}
Three-dimensional ground-state density profile for the same
parameters an in Fig.~\ref{fig4}}.
\end{center}
\end{figure*}

Figure~\ref{fig3} shows a stability diagram typical of higher
values of the aspect ratio $\lambda_y$, in this case $\lambda_y =
7$, for $0.4 < \lambda_z<1$. As $D$ is increased, the ground-state
density of the condensate first undergoes a transition from a
gaussian-like to a double-peaked profile of the type illustrated
in Figs. 4b and 5b (region III). As $D$ is further increased, this
domain is followed for $\lambda_z$ close enough to unity by a
second transition to a domain (region V) with the appearance of a
density minimum near trap center. Initially, this minimum is
surrounded by a region with a radial density modulation, see Fig.
4c, but for larger values of $\lambda_z$ this modulation is
reduced, see Figs. 4d and 5d. In that region, the density profile
resembles the solution previously reported in Ref. \cite{stabdp2}
for a similar parameter range.


In the case of atoms a typical magnetic moment of $6 \mu_B$, and
we find that the range of critical dipole strengths $D$
corresponding to structured ground states can be achieved for
$10^4-10^5$ atoms for trapping frequencies $\omega_x \approx
1kHz$.  For molecules with a typical electric dipole moment of $1
Debye$ the corresponding number is $ 10^3-10^4$ molecules. While
these are relatively high particle numbers, especially for the
atomic case, they do not seem out of reach of experimental
realization.

\section{conclusion}
In conclusion, we have performed a detailed numerical study
of the ground state structure and stability of ultracold dipolar
bosons in an anisotropic trap for dipoles polarized along the
$y$-direction. The trap aspect ratios along $y$ and $z$ direction,
$\lambda_y$ and $\lambda_z$ were used as control parameters, and
the mean-field stability diagram has established as function of
these parameters and a dimensionless interaction strength $D$. For
small $\lambda_y$ the system was found to exhibit a standard
density profile, but for larger values, and depending on
$\lambda_z$, various structured ground state were found to appear
before reaching the unstable regime where the condensate
collapses. These include a four-peak structured solution in the
$x-z$ plane, a ring-like ground state with a modulated radial
density profile. For $\lambda_y \sim 7$ and $\lambda_z=1$, we
found a biconcave condensate profile, as already reported in
\cite{stabdp2}.

For strong confining potentials along the dipole polarization
direction, i.e for large $\lambda_y$, increasing $\lambda_z$ can
be viewed as resulting in a change from a quasi-one dimensional to
a quasi-two-dimensional geometry. As such we can think of the
various ground-state structures as a result of dimensional
crossover in a trapping geometry. To gain a deeper understanding
of these structures as we approach the instability region, future
work study the Bogoliubov spectrum of the trapped system.

\acknowledgements We thank Drs. D. O'Dell, D. Meiser and R.
Kanamoto for numerous useful discussions and comments. This work
is supported in part by the US Office of Naval Research, by the
National Science Foundation, by the US Army Research Office, by
the Joint Services Optics Program, and by the National Aeronautics
and Space Administration.

\end{document}